\documentclass[sigconf ]{acmart}
\usepackage{fancyhdr}
 \usepackage{amsmath}
\usepackage{cases}
\usepackage{stfloats}
\usepackage{enumerate}
\usepackage{url}
\usepackage{graphicx}
\usepackage{balance}
\usepackage{algorithm}
\usepackage{algpseudocode}

 \usepackage{etoolbox}

\newtoggle{ACM}
\toggletrue{ACM}
\newtoggle{IEEEcls}
\toggletrue{IEEEcls}
\togglefalse{IEEEcls}

\floatname{algorithm}{Algorithm}

\iftoggle{IEEEcls}{
\usepackage{amsthm}
\theoremstyle{plain}

\theoremstyle{definition}
\newtheorem{definition}{Definition}

}

\newif\ifNotUse  %定义条件变量v1; v1true时编译，v1false时不编译%\ifnew 不支持nesting
 \NotUsetrue

 \iftoggle{ACM}{
\setcopyright{none}  % not set the copyright
\acmDOI{}
\acmPrice{}
\acmISBN{}
\acmConference[ ]{ }{ }{ }
\acmYear{2018}

}

\begin{document}
%\title{Real-time Super Hosts Detection On high-speed Network By A Novel Efficient GPU Algorithm}
\title{Regain Sliding super point from distributed edge routers by GPU}
\author{Jie Xu}
\affiliation{%
  \institution{School of computer science and engineer, Southeast university}
  %\streetaddress{8600 Datapoint Drive}
  \city{Nanjing}
  \state{China}
 % \postcode{78229}
 }
\email{xujieip@163.com}

\iftoggle{IEEEcls}{
\author{\IEEEauthorblockN{Jie Xu
                           
                                  }
\IEEEauthorblockA{School of Computer Science and Engineering\\ South East University \\
Nanjing, China\\
Email: xujieip@163.com} 
 
}

}

\begin{abstract}
Sliding super point is a special host defined under sliding time window with which there are huge other hosts contact. It plays important roles in network security and management. But how to detect them in real time from nowadays high-speed network which contains several distributed routers is a hard task. Distributed sliding super point detection requires an algorithm that can estimate the number of contacting hosts incrementally, scan packets faster than their flowing speed and reconstruct sliding super point at the end of a time period. But no existing algorithm satisfies these three requirements simultaneously. To solve this problem, this paper firstly proposed a distributed sliding super point detection algorithm running on GPU. The advantage of this algorithm comes from a novel sliding estimator, which can estimate contacting host number incrementally under a sliding window, and a set of reversible hash functions, by which sliding super points could be regained without storing additional data such as IP list. There are two main procedures in this algorithm: packets scanning and sliding super points reconstruction. Both could run parallel without any data reading conflict. When deployed on a low cost GPU, this algorithm could deal with traffic with bandwidth as high as 680 Gb/s. A real world core network traffic is used to evaluate the performance of this sliding super point detection algorithm on a cheap GPU, Nvidia GTX950 with 4 GB graphic memory. Experiments comparing with other algorithms under discrete time window show that this algorithm has the highest accuracy. Under sliding time widow, this algorithm has the same performance as in discrete time window, where no other algorithms can work.
\end{abstract} 
 
  \iftoggle{IEEEcls}{
%\keywords{ 
}


\maketitle
\section{Introduction}
With the arrival of zettabyte era\cite{cisco:NetForcast}, Internet becomes one of the most important foundations of nowadays people's life. Its health and robustness determines the development of economical, industry, education and so on. In order to let network run well, managers should monitor hosts in it and measure their state. Every second, more than 6 million packets will cross the network for a 40 Gb/s network where every packet's average size is 800 bytes \cite{expdata:Caida}. Keeping every host's state in memory from these high speed packets in real time is a heavy burden. 

But not all hosts are what managers interesting about because many of the hosts are just normal users: browsing site, watching video, sending or receiving email or something else like this. Only a small fraction of hosts, which play important roles in network, needed to be monitored specially. We call these hosts as special hosts. Special hosts could be classified by their functions, such as web server, email server, attacker. They also could be classified by traffic features, such as big traffic size host, high linking host. We can find out what application a host running by checking its packets with DPI \cite{PC2016:AHighThroughputDPIEngineOnGPUviaAlgorithmImplementationCoOptimization}. But DPI sans a packet carefully with many operations which will slow down the packets processing speed. It's not reasonable to scan every packet by DPI. Unlike DPI, traffic information of packets could be acquired by just focusing on their IP headers and traffic special host could be detected out more efficiently. All function special hosts and traffic special hosts only take a small part of hosts. Function special host is also a kind of traffic special host. So we can find traffic special hosts firstly, and scan packets of traffic special hosts by DPI to find out function special hosts. This paper researches how to detect a traffic special host, sliding super point, from the perspective of host's linking.

Super point is a host with which there are many others communicate in a time period. When the time period is sliding time window, we call the host as sliding super point. Many function special hosts are super points, for example network scanners, P2P delivers, servers, DDos attackers. Super point detection helps to locate such network events efficiently and got many researchers'  interesting\cite{HSD:2014ANewSketchMethodMeasuringHostConnectionDegreeDistribution}\cite{HSD:Infcom:ANovelDataStreamingMethodDetectingSuperpoints}\cite{HSD:samplingAdaptiveSamplingForDetectiSuperpoints}\cite{HSD:Infcom:SimpleAdaptiveIdentificationSuperspreadersFlowSampling}.

But super point is defined under discrete time window and the result will be affected by the time window's start point. What's more, super point will not be reported until the end of a time window. To overcome this problem, sliding super point is desired for monitoring hosts in a continuous and more granular way.

Sliding super point detection is more difficult than that of super point, especially in a big network which contains several distributed edge routers. To detect sliding super point in distributed environment, an algorithm must have the ability to update hosts' linking state incrementally when window sliding forward, remove stale linking state that not belong to present window, collect linking state from distributed node and regain hosts from them. For the sake of real time detection, this kind of algorithm is required to be able to run parallel. Until this paper, no such algorithm that satisfies all these requirements has ever been proposed. The deficiency of efficient detection algorithm limits the widely application of sliding super point in network field. 

This paper firstly proposed a distributed sliding super point detection algorithm which can run parallel. This algorithm just contains simple operations, no float computing operation when scan packets. It can be deployed on GPU with a little modification. GPU overcomes the two obstacles of speed: plenty computing resources and low latency of memory operation. Firstly, the great amount of cores locating on a chip let it have the ability to dealing with several packets parallel\cite{HPDC2013:AdeclarativeMulticorePlatformforScalableComposableTrafficAnalytics}\cite{HPDC2013_OnTheEfficacyGPU-integratedMPIScientificApplications}. Secondly, unlike other researches, which try to reduce memory latency by using fast but very expensive memory\cite{HPDC2016:IMPACCATightlyIntegratedMPIOpenACCFrameworkExploitingSharedMemoryParallelism}, GPU has several memory controllers which can access memory at the same time\cite{HPDC2016:ImprovingGPUPerformanceThroughResourceSharing}. And the memory latency will be concealed by those parallel launching threads\cite{HPCA2016_Warped-preexecutionAGPUpre-executionApproachImprovingLatencyHiding}. So GPU is the best platform for sliding super point detection. This paper makes the following contributions.
\begin{enumerate}
\item A novel sliding estimator is proposed in this paper. It can estimate contacting host number incrementally under a sliding window. The updating procedure of this estimator is very simple, only contains integer adding, setting and comparing operations. And it can be updated by several threads at the same time, which is very suitable for parallel and distributing environment. 

\item A reversible hash functions group is designed for regaining sliding super points. This hash functions group has high randomness which helps to make full use of every sliding estimator and save memory. What's more, a host could be reconstructed from its hashed values by this hash functions group. This makes sure that sliding super points could be regained successfully.

\item A new distributed sliding super points detection algorithm is devised based on the sliding estimator and reversible hash functions group. A single thread version and a high speed parallel version of this algorithm are proposed at the same time. We implement our algorithm on a low cost GPU to evaluate its performance in real-world traffic.

\end{enumerate}

This paper is organized by the following way. In the next section, we introduce related super points detection algorithm. Our novel sliding super points algorithm is described in detail in section 3. Section 4 shows how to implement our algorithm in GPU efficiently. Experiments on real-world core network traffic are shown in section 5. At last section we make a conclusion about this algorithm.
\section{Related work}
High speed network super point detection has been researched for a long time. At first, sampling method was used to solve the problem of slow processing speed\cite{HSD:streamingAlgorithmFastDetectionSuperspreaders}\cite{HSD:identifyHighCardinalityHosts}\cite{HSD:findFrequentItemsInStream}. But sampling method affected the accuracy of these algorithm especially in the situation where a high sampling rate was adopted. Then many works tried to improve the processing speed by using high speed memory, such as CBF\cite{HSD:LineSpeedAccurateSuperspreaderIdentificationDynamicErrorCompensation}, DCDS\cite{HSD:ADataStreamingMethodMonitorHostConnectionDegreeHighSpeed}  , VBFA\cite{HSD:DetectionSuperpointsVectorBloomFilter}.

Chen et al. \cite{HSD:LineSpeedAccurateSuperspreaderIdentificationDynamicErrorCompensation} proposed a contacting hosts estimator called counter bloom filter CBF based on the theorem of bloom filter. When a flow appears, several counters in CBF were added by one. A flow could only updated CBF once. This algorithm had a high accuracy when running with a single thread on SRAM. According to the statement of the authors, this algorithm could scan 2 million packets per second. But this speed was still too low for nowadays high speed network which forwards more than 6 million packets every seconds. And this algorithm couldn't work on parallel and distributed environment because a flow may update CBF many times in these cases. 

Wang et al.\cite{HSD:ADataStreamingMethodMonitorHostConnectionDegreeHighSpeed} used linear estimator \cite{DC:aLinearTimeProbabilisticCountingDatabaseApp} to estimate contacting hosts number and proposed a novel structure called DCDS based on Chinese Remainder Theory(CRT) which can restore hosts directly. But CRT is so complex that it requires many computing resource and time. To overcome this weakness, Liu et al.\cite{HSD:DetectionSuperpointsVectorBloomFilter} proposed an structure called VBF which was similar to a bloom filter. VBF regained hosts by bits comparing and concatenation, instead of by CRT. VBF had a much faster speed than DCDS because of its simple regaining procedure. VBF used sub bits of IP address to map a host to several linear estimators. Sub bits can be acquired quickly but had little randomness which caused that most of linear estimators in VBF were not be used and memory was wasted.

Those algorithms only focused on how to speed up by reducing memory latency and they neglected the huge computing resource requirement. GPU can solve this two problem, high memory operation speed and plenty computing resource, all together.  

GPU is the best desktop super computing platform which has the same computing ability as a small cluster. In a single GPU chip, hundreds or thousands of cores sharing a big global graphic memory. Different threads can read and store this memory parallel. Although a core in GPU is a little slower, lower frequency, than a core in CPU, the total computing ability of these hundreds of GPU cores is much stronger than that of a CPU which only have teens of cores at most. The convenient program environment, such as CUDA\cite{GPU:OptimizationPrinciplesApplicationPerformanceCUDA}, OpenCL\cite{OpenCL:AParallelProgrammingStandardForHeterogeneousComputingSystems}, let GPU become one of the most popular parallel computing platform. 

GPU was firstly used to detect super points by Seon-Ho et al.\cite{HSD:GPU:2014:AGrandSpreadEstimatorUsingGPU}. They deployed a novel structure called virtual vector on GPU to estimate contacting hosts. But virtual vector can only estimate contacting hosts number, super points can't be reconstructed from it directly.

All of these works can only detect super points and each of them have their own limitations. This paper will introduce a sliding super points detection algorithm and introduce how to deploy it on GPU for real time distributed running.

\section{Distributed Slinding super point detection}
\subsection{Sliding super point}
\begin{figure}[!ht]
\centering
\includegraphics[width=0.47\textwidth]{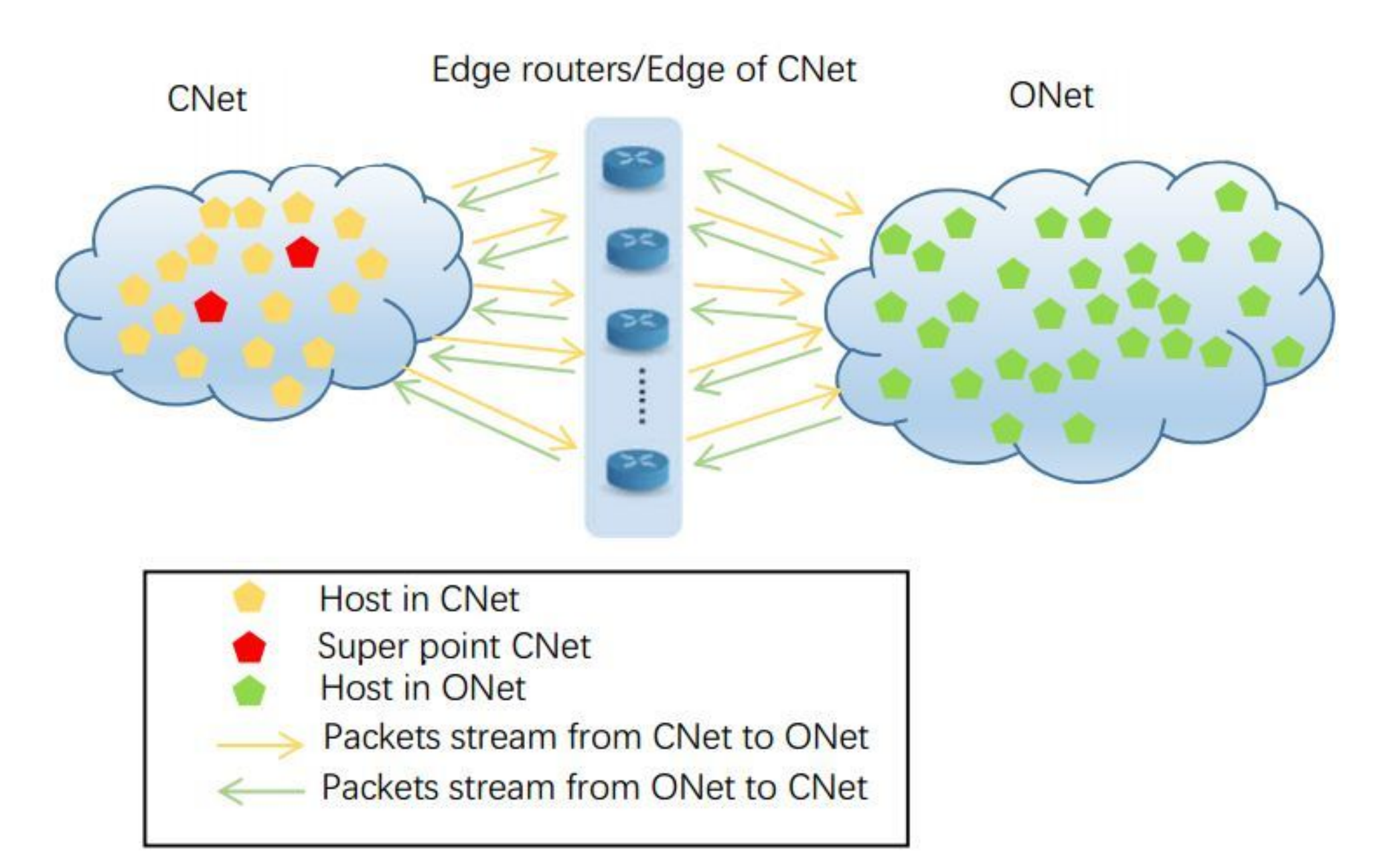}
\caption{Packets stream between two networks}
\label{NetworkEdgeRouters}
\end{figure}

Suppose there is a core network $CNet$ under managing of some ISP(internet service provider) or organization. It communicates with another network $ONet$ through a set of edge routers $RS$ as shown in figure \ref{NetworkEdgeRouters}. $RS$ is although called the edge of $CNet$ and $ONet$. 

Host in this paper means a device or a virtual machine with a unique IP addresses and it can be represented by its IP address. If a host $cip$ in $CNet$ wants to contact with another host $oip$ in $ONet$, $RS$ is required to forward their packets. At $RS$, two direction packets stream could be observed: from $CNet$ to $ONet$ and from $ONet$ to $CNet$. A IP address pair stream like $\{<cip_0,oip_0>, <cip_2,oip_1>,<cip_i,oip_j>,\cdots\}$ could be extracted from the two packet streams where $cip_i \in CNet$ and $oip_j \in ONet$. Sliding super point will be acquired from the IP pair stream. 

For a manager of $CNet$, he wants to monitor the running state of $CNet$ and keep its security. So hosts in $CNet$ is under particular scrutiny and the following definitions and algorithms are focus on $CNet$. Situation of $ONet$ could be derived by change the position of $CNet$ and $ONet$. From the perspective of $RS$, we give the following definition. 

\begin{definition}[Opposite Points]
\label{def-cnet_oppositePoints}
For a host $cip$ in $CNet$, the set of hosts in $ONet$ that send packets to it or receive packets from it in a certain time period $t$ is  $cip$'s opposite points, written as $OP(cip,t)$.
The number of host in $OP(cip,t)$ is called its opposite number, written as $|OP(cip,t)|$.
\end{definition}

\begin{definition}[Super Point]
\label{def-superPoint}
For a host $cip$ in $CNet$, if its opposite number $|OP(cip,t)|$ in time period $t$ is no less than a certain threshold $\theta$, $cip$ is a super point in this time period.
\end{definition}

When the time period in definition \ref{def-superPoint} is a sliding time window, the super point is called sliding super point.

\begin{definition}[Sliding time window]
\label{def-slidingTimeWindow}
For two time point $sartT$ and $endT$, the duration $endT - startT$ between them is divided into n successive slots $\{s_0,s_1,s_2,s_3,\cdots,s_{n-1}\}$ where each slot $s_i$ has the same duration $\mu = \frac{endT - startT}{n}$. A sliding window is a time period composed of $k$ successive slots starting from slot $s_i$, denoted as $SW(i,k)$.
\end{definition}

\begin{figure}[!ht]
\centering
\includegraphics[width=0.47\textwidth]{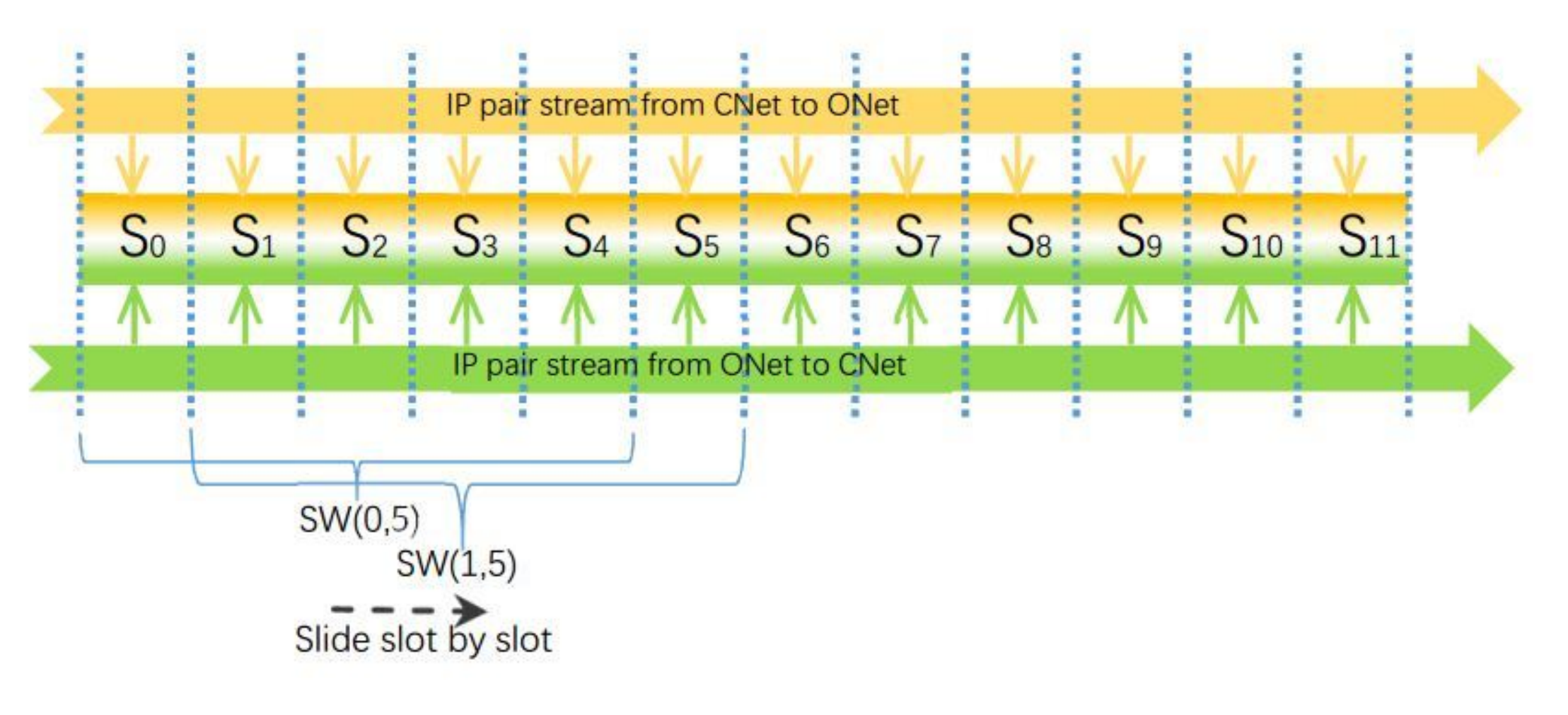}
\caption{IP pair stream under sliding window}
\label{SlidingSlotsTrafficPartition}
\end{figure}

There are $k-1$ slots are the same in two adjacent sliding windows $SW(i,k)$ and $SW(i+1,k)$ because sliding window moves forward one slot at a time. The traditional time window, called as discrete window, is a special case of sliding window that $k=1$. There is no overlapped time slice in different discrete windows. For a discrete window with duration $k*\mu$, it will move forward $k$ slots at a time instead of one. So discrete window can't give a continual observation duration like sliding window. IP pair streams could be parted into successive slot logically like figure \ref{SlidingSlotsTrafficPartition}. In figure \ref{SlidingSlotsTrafficPartition}, $k$ is set to 5 and IP pair streams contains all IP pairs extracted from both directions IP packet stream.

Super point under sliding window is called sliding super point.

\begin{definition}[Sliding Super Point, SSP]
\label{def-SlidingSuperPoint}
Suppose a traffic is divided into successive slots and each slot has duration $\mu$. For a host $cip \in CNet$, if $|OP(cip,SW(i,k))| \geq \theta$ from slot $S_i$ to $S_{i+k-1}$, then $cip$ is a sliding super point. The set of sliding super points at sliding window $SW(i,k)$ is written as $SSP(i,k)$.
\end{definition}

The task of sliding super point detection is to find out $SSP(i,k)$. SSP provides more granularity for host monitor. Taking traffic in figure \ref{SlidingSlotsTrafficPartition} as an example, if there are $\theta$ hosts in $ONet$ communicating with a host $cip$ in $CNet$ in the six slot $S_5$ and there are no other hosts in $ONet$ communicate with $cip$ in other slots. Under sliding window, $cip$ will be detected out at $SSP(1,5)$, $SSP(2,5)$, $SSP(3,5)$, $SSP(4,5)$, $SSP(5,5)$. But under discrete window which has the same duration $k*\mu$ and starts from the begin of first slot $S_0$, $cip$ will be reported only in the second window last from $S_5$ to $S_9$. What's more, $cip$ will be reported as soon as the slot $S_5$ is finished under sliding window. While under discrete window, $cip$ won't be detected out until the end of $S_9$. 

Although SSP monitors hosts more precisely, it puts forwards higher requirements for opposite number calculation. 

\subsection{Opposite number estimation under sliding window}
Opposite number calculation is the foundation of super point or sliding super point detection. We firstly discuss the situation under discrete window. One of the simplest ways of opposite number calculation is keeping every appearing opposite host in memory and adding a new host to the set when coming a IP pair whose opposite host is not in memory. A host's opposite number is the number of hosts keeping in memory. We call this method as precise algorithm. 

For example, suppose $cip \in CNet$ and there are four IP pairs contacting with it in a time period $t$: $\{IPpair_1=<cip, oip_1>, IPpair_2=<cip, oip_1>, IPpair_3=<cip, oip_2>, IPpair_4=<cip, oip_1>\}$. By precise algorithm, a big enough memory buffer is allocated before scanning IP pairs. When scanning $IPpair_1$, $oip_1$ will be checked in the buffer. Because there are no hosts in the buffer at the begin, so $oip_1$ will be inserted to it. $oip_1$ is in $IPpair_2$ too and it will be found appearing in the buffer. So $oip_1$ won't be added to the buffer when scanning $IPpair_2$. When scanning $IPpair_3$, precise algorithm looks for $oip_2$ in the buffer and inserts it to the buffer because not finding. No host will be inserted into the buffer when scanning $IPpair_4$ because $oip_1$ will be found already appearing in the buffer. At the end of $t$, there are two hosts in the buffer totally and $|cip|=2$.

The merit of precise algorithm is that it has the absolute accuracy. For a host $cip$ in $CNet$, $OP(cip)$ will be constructed at the end of a time period by precisely algorithm. But precise algorithm has two drawbacks which limit its application in high speed network: high memory requirement and low processing speed.

Precise algorithm stores every opposite host in memory. For a host $cip$ under DDos flooding attacks, there are huge faked hosts in $OP(cip)$ requiring lots of memory to keep them. When $CNet$ is a core network, it will contain huge hosts too. Allocating a buffer for every host in $CNet$ is a burden for memory management. 

Precise algorithm requires many memory access and can't scan packets parallel, which prevent its speed raising. When scanning a IP pair, precise algorithm needs to know if the opposite IP address has already appeared by querying the buffer. This checking procedure contains many memory operation and its accuracy depends on the buffer coherence. In the previous sample, if there are four threads scanning these four IP pairs separately and parallel, one thread deals with a IP pair. These four threads will query the buff at the same time. Because there is no host in the buffer, so every opposite IP address will be inserted to it. In this situation, the buffer contains three $oip_1$ and one $oip_2$. $|OP(cip)|$ is regarded as to be 4 by mistake. Without parallel running, precise algorithm can't scan high speed packets in real time. It always be used to acquire the accuracy answer offline, as the baseline to judge other algorithms' accuracy.

To overcome the weakness of precise algorithm, estimating algorithm is widely used in high speed network super point detection. It has contrary features against precise algorithm, fixed and small memory requirement, parallel ability, little deviation.

Linear estimator \cite{DC:aLinearTimeProbabilisticCountingDatabaseApp} is one of the best estimating algorithms. Small memory occupation, high accuracy and simple updating procedure, linear estimator got many researchers' attention. Linear estimator uses $\eta$ bits, initialized with 0, to record opposite hosts appearance.  When a IP pair with opposite host $oip_0$ appears, linear estimator will choose and set a random bit, determined by a hash function $H_1(oip_0)$. $H_1$ is a hash function \cite{hash_ThePowerOfSimpleTabulationHashing} which maps a IPv4 address to a random value between 0 and $\eta -1$. At the end of a time period, opposite number could be evaluated by counting the remaining zero bit number $z_0$ according equation \ref{eqt_LDC_estValueFromZeroBitsN}.
\begin{equation}\label{eqt_LDC_estValueFromZeroBitsN}
 {Est}'=- \eta *ln(\frac{z_0}{\eta })
\end{equation}

Although linear estimator has an excellent performance under discrete window, it can't be applied to sliding window because it does not keep the opposite hosts information of previous time period. After estimating opposite number, linear estimator will reinitialize every bit to zero for next time window. But in sliding window $SW(i,k)$, we focus on hosts not only in the scanning slots but also its previous (k-1) slots. How to remove hosts not in $SW(i,k)$ exactly is the key in sliding opposite number estimating. 

A novel estimator, called as sliding estimator, is devised in this paper for opposite number estimating under sliding window. Unlike linear estimator, sliding estimator keeps the state of hosts appearing in the previous (k-1) slots. 

Sliding window uses short integer array instead of bit array using in linear estimator to record the appearing of a host. There are $\eta$ short integers in sliding estimator and each short integer occupies 2 bytes. Short integer in sliding estimator is used to record the distance of the nearest slot that mapped by an opposite host from now scanning slot. So every short integer in sliding estimator is also called as distance recorder. Every distance recorder should be initialized before scanning IP pairs. Unlike linear estimator, the initialize operation only need once at the begin of algorithm, not before every slot. Because a short integer (short integer in this paper means unsigned short integer) can reach 65535 at most, the nearest slot distance in sliding window could be 65535 and k must smaller than 65535. But this is long enough for most monitoring task. When k is set to 65534 and $\mu$ is set to 1 second, a sliding window is as long as 18.2 hours. Every distance recorder is initialized to its biggest value 65535 before scanning IP pair. For a sling estimator $SE$, let $SE[i]$ point to its $i$th distance recorder. When updating an opposite host $oip_0$ to $SE$, $SE[H_1(oip_1)]$ is set to 0. This is because the distance of a slot to itself is 0. 

Sliding estimator records all opposite hosts and their appearing slots. At the end of a slot, we should calculate the number of distance recorders that being updated within k slots. This number, written as $R_k$, could be acquired by counting distance recorders whose values are littler than k. $R_k$ has the same meaning as $\eta - z_0$ in discrete window whose size is $k*\mu$. So opposite number in sliding window could be estimated by equation \ref{eqt_LDC_estValueFromZeroBitsN} with $z_0= \eta - R_k$.

\begin{figure}[!ht]
\centering
\includegraphics[width=0.47\textwidth]{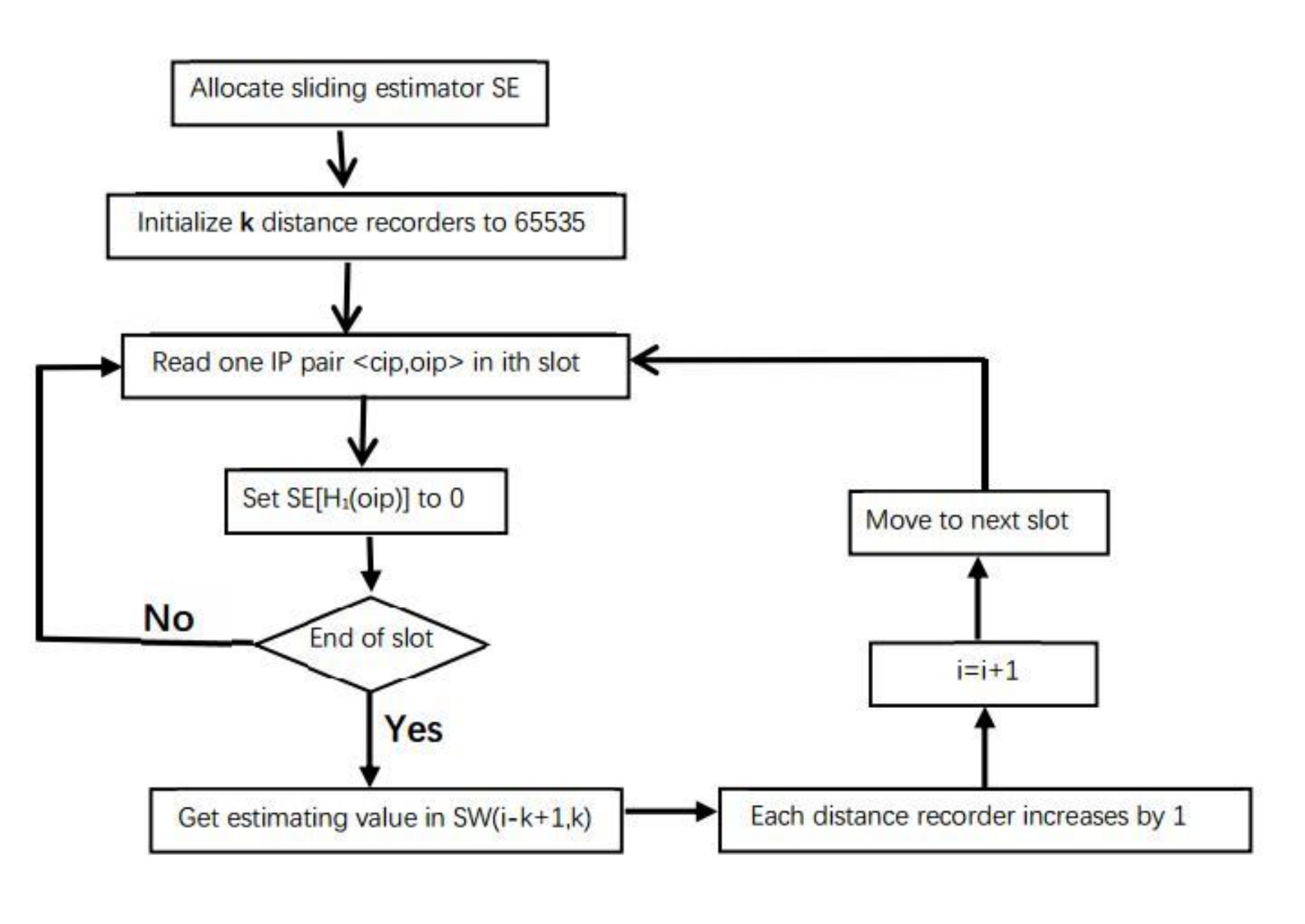}
\caption{Sliding estimator procedure}
\label{SlidingEstimatingProcedure}
\end{figure}

When sliding window $SW(i,k)$ moves one slot to $SW(i+1,k)$, distance of nearest slot will increase too. So at the beginning of every slot, each distance recorder will be added by 1 if its value is littler than 65535. Figure \ref{SlidingEstimatingProcedure} shows how sliding estimator works.

Figure \ref{SlidingEstimatingProcedure} shows that, once a sliding estimator $SE$ is allocated, it can work all along with the sliding window moving forward. For every IP pair in a slot, sliding estimator deals with it by simply set a distance recorder to zero. This simple operation let sliding estimator have a fast IP pair scanning speed. Like linear estimator, sliding estimator uses fixed size of memory and can scan several IP pairs parallel. 

Besides the operation displaying in figure \ref{SlidingEstimatingProcedure}, there is another process relating with sliding estimator: sliding estimator merging. Several sliding estimators could be merged into a new one as describing in algorithm \ref{unionSE}. This merging process is widely used in the follow sections, sliding super points detection and distributed nodes merging. 
\begin{algorithm}                       
\caption{UnionSE}          
\label{unionSE}                            
\begin{algorithmic}                    
\Require {\\sliding estimator set $SEset$,\\
Distance recorders number $\eta$} 
\Ensure{\\ union slinding estimator $USE$\\}  
\State $USE \Leftarrow $ new sliding window  
\For{$i \in [0,\eta -1]$}
\State $maxV \Leftarrow 0$
\For{ $se \in SEset$}
\If {$maxV < se[i]$}
\State $maxV \Leftarrow se[i]$
\EndIf
\EndFor 
\State $USE[i] \Leftarrow maxV$
\EndFor
\State Return $USE$
\end{algorithmic}
\end{algorithm}

Sliding estimator will be used to estimate opposite number of different hosts in the sliding super point detection. But it is too expensive and low efficient to allocate a sliding estimator for every host due to the huge number of hosts in a core network. A novel structure consists of fix number of sliding estimators is devised for sliding super point detection. The next part discusses how to mining sliding super point by this structure in detail.
\subsection{Sliding super point detection}
Based on sliding estimator, a smart structure, called as reversible sliding estimator array RSEA, is proposed. RSEA contains $2^q$ columns and r rows of sliding estimators. Let RSEA[i,j] point to the sliding estimator in the $i$th row, $j$th column. This structure is reversible because sliding super point could be reconstructed from it without any other data. This reversible ability comes from a novel hash functions group, reversible hash functions group RHFG. 

RHFG is an array of r hash functions, each of which hashes an IP address to a value between 0 and $2^q-1$. Let RHFG[i] represent the $i$th hash function. RHFG[0] is a random hash function \cite{hash_ThePowerOfSimpleTabulationHashing} which maps a IP address to an integer between 0 and $2^q-1$, $0 \leq RHFG[0](cip) \leq 2^q-1$ where $cip \in CNet$. The rest $r-1$ hash functions are derived from $RHFG[0]$ according the following equation.
\begin{equation}\label{eqt_RHFG_restHashes}
RHFG[i](cip)= ((cip>> (i*\delta) ) XOR RHFG[0](cip) ) mod (2^q)
\end{equation}

In equation \ref{eqt_RHFG_restHashes}, $1 \leq i \leq r-1$, $cip \in CNet$. $\delta$ is an positive integer that smaller than q and $(r-2)*\delta+q \geq 32$. "XOR" is the bit wise exclusive or operation. "$>>$" is the bit wise right shift operation. According to the property of "XOR", we can recover $(cip>> (i*\delta)) mod (2^q)$ by equation \ref{eqt_RHFG_bits_from_xor_cid}.

\begin{equation}\label{eqt_RHFG_bits_from_xor_cid}
(cip>> (i*\delta) ) mod (2^q)= RHFG[i](cip) XOR RHFG[0](cip)  
\end{equation}

$(cip>> (i*\delta) ) mod (2^q)$ is q successive bits of $cip$ starting from $i*\delta$, written as $B(i)$. Because $(r-2)*\delta+q \geq 32$, every bit in $cip$ will appear in some $B(i)$ where $1 \leq i \leq r-1$. For a host $cip \in CNet$, let RHFG(cip) represent the array of r hashed values where $RHFG(cip)=\{RHFG[0](cip),RHFG[1](cip),\cdots,RHFG[r-1](cip)\}$. $cip$ could be regained from RHFG(cip) by extracting bits from $B(i)$.

RHFG has high randomness and reversible ability. It is used to select $r$ sliding estimators from each row of RSEA for every host in $CNet$. For a host $cip$ in $CNet$, its r sliding estimators are denoted as $RSEA(cip)=\{RSEA[0,RHFG[0](cip)], RSEA[1,RHFG[1](cip)],\cdots,\\ RSEA[r-1,RHFG[r-1](cip)]\}$. When a IP pair comes, these r sliding estimators will be updated at the same time as shown in algorithm \ref{alg_UpdateRSEA_oneIPpair}.

\begin{algorithm}                       
\caption{Update RSEA}          
\label{alg_UpdateRSEA_oneIPpair}                            
\begin{algorithmic}                    
\Require {\\IP pair $<cip,oip>$,\\
Reversible hash functions group $RHFG$,\\
Reversible sliding estimator array $RSEA$\\ } 
\State $DRidx \Leftarrow H_1(oip)$  
\For{$i \in [0,r -1]$}
\State $COLidx \Leftarrow RHFG[i](cip)$ 
\State $se \Leftarrow RSEA[i, COLidx]$
\State $se[DRidx] \Leftarrow 0$ \label{alg-line-updateRSEA_SEupdate}
\EndFor 
\end{algorithmic}
\end{algorithm}

For every IP pair, r distance recorders in RSEA will be set to 0 in algorithm \ref{alg_UpdateRSEA_oneIPpair}. These operations could be done parallel without any conflicts. After scanning all IP pairs in a slot, sliding super point will be reconstructed from RSEA. According to the feature of RHFG, if $RSEA(cip)$ is known, $cip$ could be restored from it. But in the simple IP pair updating process, $RSEA(cip)$ is not stored directly. According to the definition, if $cip$ is a sliding super point, every sliding estimator in $RSEA(cip)$ will contain no less than $\theta$ opposite hosts. The sliding estimator in whose opposite number is no less than $\theta$ is called as hot sliding estimator denoted by $HSE$. According to equation \ref{eqt_LDC_estValueFromZeroBitsN}, there is a threshold $\widehat{R_k}$ to judge if a sliding estimator is a $HSE$, $\widehat{R_k}=\eta*(1-e^{- \frac{\theta}{\eta}})$. Only when a sliding estimator's $R_k$ is equal to or bigger than $\widehat{R_k}$ will it be judged as a $HSE$. Let $HSE(i)$ mean the set of $HSE$ in the $i$th row and $|HSE(i)|$ be the number of elements in $HSE(i)$. 
A candidate tuple $CT$ consists of r $HSE$ could be acquired by selecting a $HSE$ from every $HSE(i)$ where $0 \leq i \leq r-1$. $CT=\{he_0,he_1,he_2, \cdots ,he_{r-1}\}$ where $he_i \in HE(i)$. Sliding super point could be regained by testing all of these candidate tuples. 

Notice that, every $B(i)$ in equation \ref{eqt_RHFG_bits_from_xor_cid} has a feature that its left $q-\delta$ bits is equal to the right $q-\delta$ bits of $B(i+1)$ where $1 \leq i \leq r-2$. So not every candidate tuple can reconstruct a valid IP address. By this property, sliding super point could be restored incrementally from $HSE(1)$ to $HSE(r-1)$ as describing in algorithm \ref{alg-restorSSP_fromHSE}.

\begin{algorithm}                       
\caption{Regain sliding super point}          
\label{alg-restorSSP_fromHSE}                            
\begin{algorithmic}                    
\Require {\\  %Reversible hash functions group $RHFG$,\\
Reversible sliding estimator array $RSEA$} 
\Ensure {\\ Sliding super point list $SSPL$\\}
\For{$i \in [0,2]$}
\State $HSE(i) \Leftarrow $ HSE in $i$th row of $RSEA$ 
\EndFor 
\For{ $CT=<he_0,he_1,he_2> in <HE(0),HE(1),HE(2)>$}
\State $B(1) \Leftarrow he_0 XOR he_1$
\State $B(2) \Leftarrow he_0 XOR he_2$
\If{ left $q-\delta$ bits of $B(0)$ not equal to right $q-\delta$ bits of $B(1)$} \label{alg-line_BitVectorCondition}
\State Continue
\EndIf
\State $tmpSSPL \Leftarrow IRSSP(RSEA,CT_3,3)$ \label{alg-restorSSP_fromHSE_recursivellyRestor}
\State insert $tmpSSPL$ into $SSPL$
\EndFor
\end{algorithmic}
\end{algorithm}

Algorithm \ref{alg-restorSSP_fromHSE} gets HSE in the first three rows of RSEA by comparing $R_k$ of every SE with $\widehat{R_k}$ . Then check every candidate tuple extracting from $HE(0)$, $HE(1)$ and $HE(2)$. $B(1)$ and $B(2)$ derived from a candidate tuple are used to determine if there is need to check this candidate tuple further. When $B(1)$ and $B(2)$ conform to the condition in line \ref{alg-line_BitVectorCondition}, this candidate tuple maybe a part of a sliding super point's relating sliding estimators. In line \ref{alg-restorSSP_fromHSE_recursivellyRestor}, this tuple will be checked with hot estimators in other rows by function $IRSSP$, which is a recursive function as described in algorithm \ref{alg-restorSSP_scanHSE_recursively}.

\begin{algorithm}                       
\caption{Incrementally Restore SSP, IRSSP}          
\label{alg-restorSSP_scanHSE_recursively}                            
\begin{algorithmic}                    
\Require {\\  %Reversible hash functions group $RHFG$,\\
Reversible sliding estimator array $RSEA$, \\
Candidate tuple $CT_i$,\\
Row index $i$
} 
\Ensure {\\ Sliding super point list $SSPL$ \\}
\State sliding super point list $SSPL \Leftarrow NULL$
\If { number of element in $CT_i$ smaller than $i$}
\State Return $SSPL$
\EndIf
\State $HSE(i) \Leftarrow $ HSE in $i$th row of $RSEA$ 
\For{$he_i \in HSE(i)$}
\State $he_0 \Leftarrow$ first element in $CT_i$  \label{alg_line_rCheck_start}
\State $he_{i-1} \Leftarrow$ last element in $CT_i$
\State $B(i-1) \Leftarrow he_0 XOR he_{i-1}$
\State $B(i) \Leftarrow he_0 XOR he_i$
\If{ left $q-\delta$ bits of $B(0)$ not equal to right $q-\delta$ bits of $B(1)$}
\State Continue
\EndIf                                      \label{alg_line_rCheck_end}
\State $CT_{i+1} \Leftarrow CT_i$
\State push $he_i$ into $CT_{i+1}$
\If{$ i == r-2$}
% now test the last HE, 
\State $USE \Leftarrow UnionSE(CT_{i+1})$  \label{alg-line-unionSE}
\State $R_k \Leftarrow $ the number of distance recorder in USE whose value is smaller than k
\If {$R_k \geq \widehat{R_k} $}
\State extract host $cip$ from $CT_{i+1}$   \label{alg-line-regainHost}
\State insert $cip$ into $SSPL$
\EndIf
\Else
\State $tmpSSPL \Leftarrow IRSSP(RSEA,CT_{i+1},i+1)$
\State insert $tmpSSPL$ into $SSPL$
\EndIf
\EndFor
\State Return $SSPL$
\end{algorithmic}
\end{algorithm}

Algorithm \ref{alg-restorSSP_scanHSE_recursively} checks a tuple recursively with hot estimators in different rows. A candidate tuple will grow incrementally until meeting a hot estimator in the last row. When testing candidate tuple with a hot estimator in $HE(i)$, lines from \ref{alg_line_rCheck_start} to \ref{alg_line_rCheck_end} check if this hot estimator can be inserted into this candidate tuple. If it can, a new candidate tuple will be generated from this candidate tuple by adding it.  If the added hot estimator is one in the last row, a candidate host will be extracted from this new candidate tuple. A union sliding estimator derived from hot estimators in this candidate tuple in line \ref{alg-line-unionSE} will be used to check if the reconstructed host is a really sliding super point and add it to sliding super point list if it is. Host's IP is reconstructed in line \ref{alg-line-regainHost} by extracting bits in different $B(i)$ where $1 \leq i \leq r-1$. If the added hot estimator is not in the last row, the new candidate tuple will be examined with hot estimators in the next row recursively by this algorithm.

Algorithm \ref{alg-restorSSP_fromHSE} shows how to regain sliding super point in a recursive way. This way can work with only allocating a candidate tuple containing $r$ sliding estimators' indexes. It requires little memory. But this is not the fast method when running parallel in GPU because which hot estimators can generate a valid tuple is unknown and tasks can't be arranged balance. Next section introduces a fast parallel algorithm to reconstruct sliding super point making full use of GPU's memory and computing ability.

\section{Detect Sliding super point on GPU}
There are three essential procedures in our algorithm: IP pair scanning, sliding super points reconstruction, window sliding. All of them could run parallel on GPU after some modification. Our algorithm can be easily deployed on distributed GPU node. Suppose there are $n$ edger routers. Each router mirrors its traffic to a monitor server for sliding super point detection. On the monitor server, a GPU card is connected with it through PCI-E 3.0. And on the global memory of every GPU, a $RSEA$ is allocated. All of these $RSEA$ have the same rows and columns. For IP pair scanning and slots updateing, every node will process packets passing through its edge routers and update their own $RSEA$ separately. After scanning all packets of a slot, $RSEA$ on $n$ distributed monitor servers will be sent to a certain monitor server and merged into a global $RSEA$ by sliding estimator union---algorithm \ref{unionSE}. Suppose $RSEA_i$ represents the $RSEA$ on the $i$th node and $GRSEA$ represents the merged $RSEA$. Algorithm \ref{alg_mergeDistributedRSEA} illustrates how to generate $GRSEA$ from distributed $RSEA$.
\begin{algorithm}                       
\caption{merge $RSEA$ }          
\label{alg_mergeDistributedRSEA}
\begin{algorithmic}                    
\Require {\\  distributed nodes'RSEA set\\ 
             \ \ \ \ \ $RSEAS=\{RSEA_0, RSEA_1,\cdots,RSEA_{n-1}\}$
} 
     \Ensure {\\ global RSEA $GRSEA$ \\ }
\For { $rIDX \in [0,r-1]$}
\For{ $cIDX \in [0, 2^q -1]$}
   \State $RSet \Leftarrow Null$
    \For{$ i \in [0,n-1]$}
     \State insert $RSEA_i[rIDX,cIDX]$ into $RSet$
\EndFor
  $GRSEA[Ridx,cIDX] \Leftarrow UnionSE(RSet)$ \label{alg-line-setGRSEA}
\EndFor
\EndFor
\State Return $GRSEA$ 
\end{algorithmic}
\end{algorithm}
In algorithm \ref{alg_mergeDistributedRSEA}, only line \ref{alg-line-setGRSEA} writes global memory, others just read. So thousands of threads could be launched parallel to finish the task and every thread can merge a few $SE$s of $GRSEA$ to speed up this procedure. In the following, IP pair scanning and slots updating are run on distributed nodes and sliding super points are regained from global $RSEA$. In the parallel hosts reconstruction part, $RSEA$ represents $GRSEA$ simply.

\subsection{Parallel IP pair scanning}
Algorithm \ref{alg_UpdateRSEA_oneIPpair} describes how to update RSEA for a IP pair. But there are millions of IP pairs every second for example in a 40 Gb/s network. Dealing with these IP pairs one by one will consume much time for a single thread. In algorithm \ref{alg_UpdateRSEA_oneIPpair} only line \ref{alg-line-updateRSEA_SEupdate} update memory, others are computing operations such as getting sliding estimator index in RSEA, calculating which distance recorder to be set. It updates memory fewer times than precise algorithm does. A distance recorder could be set to zero multi times which makes sure that there is no need to synchronize among memory access and several IP pairs could update RSEA at the same time. 

Nowadays CPU contains several cores, from 2 to 22 or more such as Intel E5-2699v4. When exploiting all cores of CPU to scan IP pairs parallel, the processing speed will be raised. But the memory bandwidth of CPU will limit the increment. What's more, the price of CPU grows rapidly with the number of cores because the single core of CPU is so powerful that it occupies much space on chip. 

Unlike CUP's core, each core of GPU is a little simpler, lower frequency and fewer controlling unit, but occupies much smaller space. So a GPU could contain hundreds or even thousands of cores in a chip easily.  The total computation ability of GPU is much stronger than that of CPU. And GPU has a lower memory access latency because it has several memory controllers for multi threads. For tasks, which deal with different data by the same instructions, GPU can acquire a high speed-up. IP pair scanning is such a task.

\begin{figure}[!ht]
\centering
\includegraphics[width=0.47\textwidth]{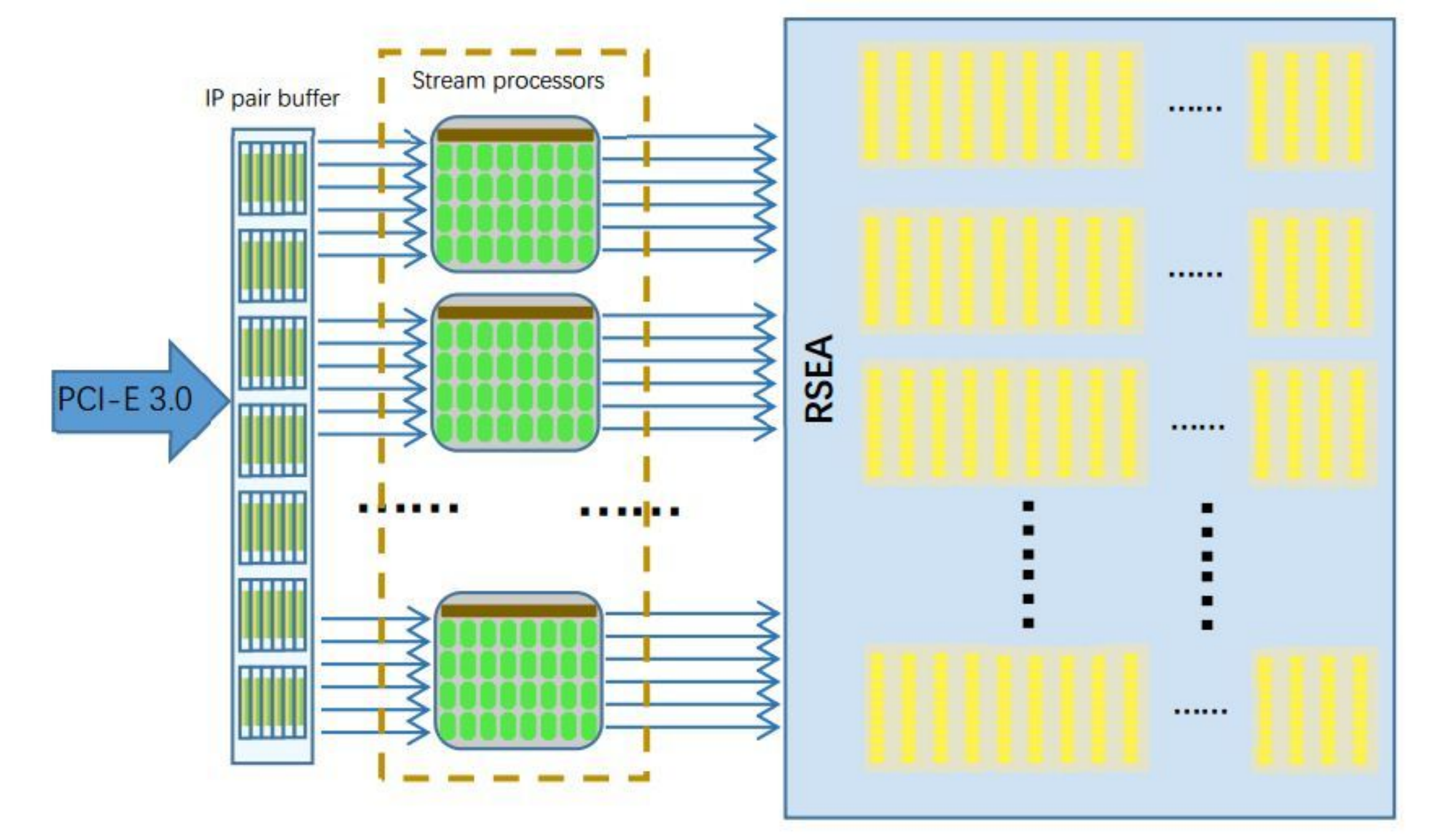}
\caption{IP pair scan on GPU}
\label{GPU_IPpairScan}
\end{figure}

IP pair scanning consumes the most time in sliding super points detection because the huge number of IP pairs appearing in every slot. Every IP pair is processed by the same algorithm, algorithm \ref{alg_UpdateRSEA_oneIPpair}. So thousands of threads running algorithm \ref{alg_UpdateRSEA_oneIPpair} could be launched to scan thousands of IP pairs at the same time. Figure \ref{GPU_IPpairScan} illustrates how to detect sliding super point on GPU.

IP pair will be copied to GPU's global memory by PCI-E bus.  A IP pair buffer on GPU memory, which can contain $\alpha$ IP pairs, is allocated to receive IP pairs. When the buffer is full or IP pairs in a slot are all copied, the same number of threads, as the number of receiving IP pairs, will be launched on GPU to process these IP pairs. Every thread reads one IP pair from global memory and updates $r$ distance recorders in RSEA which locating in global memory too. For IP v4 address, the buffer of IP pair occupies $8*\alpha$ bytes. When $\alpha$ is set to $2^{15}$, this buffer need 256 KB memory. The graphic memory on GPU, ranging from 1 GB to 11 GB, is bigger enough to hold it. Although the RSEA requires more memory than IP pair buffer, the global memory is plenty enough to store a RSEA which is big enough for a 40 Gb/s networks. Other running parameters, such as hash function parameters, r, q and $\delta$, are stored in constant memory which is read only but has high speed. A low cost GPU, which can be brought within 200 dollars, is fast enough to scan IP pairs in a 40 Gb/s network in real time.

\subsection{Parallel hosts reconstruction}
Algorithm \ref{alg-restorSSP_fromHSE} gives a recursive method to regain sliding super points. It is a memory efficient way when running on a single thread. But it is low efficient on GPU because different threads have different workload. In order to let every thread have the same scale of task a GPU sliding super point reconstructing algorithm is designed. Two additional buffers of candidate tuple are used in this algorithm: one for storing and the other for reading. Their roles exchange in different levels, when adding hot estimators in different rows. Let $SCTB$ point to the candidate tuple buffer for storing and $RCTB$ point to the candidate tuple buffer for reading.

\begin{figure}[!ht]
\centering
\includegraphics[width=0.47\textwidth]{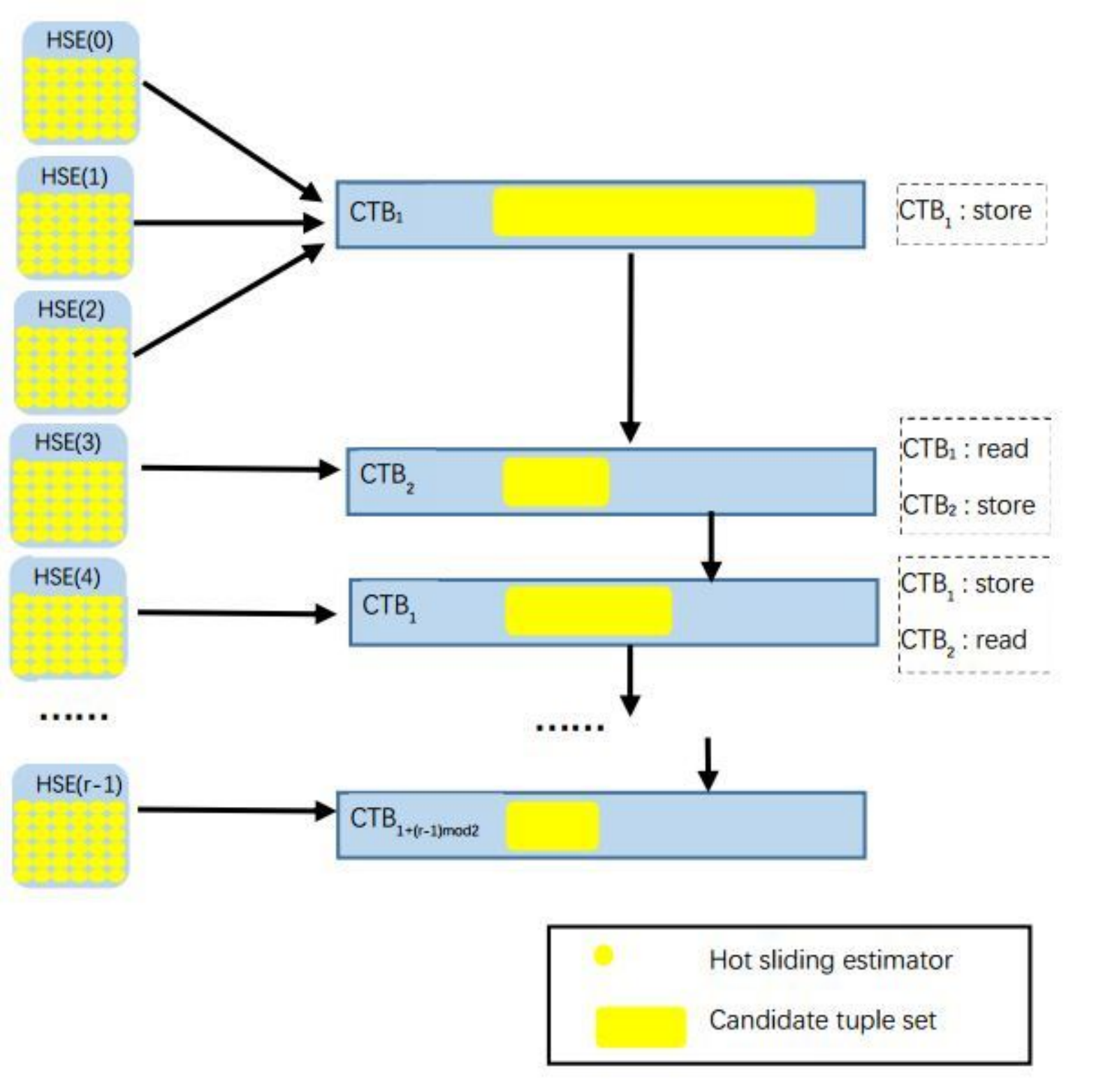}
\caption{Regain sliding super points on GPU}
\label{GPU_candidateTupleGrow}
\end{figure}

Candidate tuple in these two buffers grows incrementally from empty to a valid tuple containing $r$ hot estimators in different rows. Let $CTB_1$ and $CTB_2$ represent these two buffers respectively. Figure \ref{GPU_candidateTupleGrow} shows how candidate tuple grows with two buffers' support.

These two candidate tuple buffers are located on GPU's global memory. Candidate tuple $CT_2=\{he_0,he_1,he_2\}$ with three hot estimators, selected from $HSE(0)$ $HSE(1)$ $HSE(2)$ separately, will be inserted into $CTB_1$ after checking. The checking procedure is to test if $B(1)$ and $B(2)$ extracted from $CT_2$ is valid. Only when passing the test, will $CT_2$ be added to $CTB_1$.  When $q-\delta$ is big, only a small part of such candidate tuple will appear in $CTB_1$. The memory updating latency caused by candidate tuple insertion will be concealed by the huge parallel running threads on GPU. So candidate tuple checking determines the time consumption of a thread. When every thread deals with the same amount of candidate tuples, they will finish approximately at the same time. In this situation, the load of every thread is balance and GPU realizes its full potential.

There are total $Q=|HSE(0)|* |HSE(1)|* |HSE(2)|$ candidate tuples like $CT_2$. Suppose $V$ threads are launched on GPU to deal with these candidate tuples. Let $U$, $V$, $Q$ and $W$ be non-negative integers. In order to let every thread has the same candidate tuples to check, each thread will be assigned at least $U=Q/V$ candidate tuples evenly. Still there are $W=Q mode V$ candidate tuples rest. In these $V$ threads, every of first $W$ threads has $U+1$ candidate tuple and every of the rest $V-W$ threads has $U$ candidate tuple. Let $CT_2(i)$ represent the set of candidate tuples to be tested by the $i$th threads in GPU which can be acquired from $HSE(0)$ $HSE(1)$ $HSE(2)$. Algorithm \ref{alg_candidateBuf_H012} shows how every thread checks candidate tuples. 

\begin{algorithm}                       
\caption{global function on GPU\\
          Generate candidate tuple $CT_2$}          
\label{alg_candidateBuf_H012}
\begin{algorithmic}                    
\Require {\\  %Reversible hash functions group $RHFG$,\\
Hot sliding estimator set $HSE(0)$, $HSE(1)$, $HSE(2)$, \\
Storing candidate tuple buffer $SCTB$\\
} 
 
\State $TID \Leftarrow $ thread index 
\State get candidate tuple set from $HSE(0)$, $HSE(1)$, $HSE(2)$
\State $CT_2(TID) \Leftarrow$ candidate tuple set to be tested by this thread
 
\For{$ct=\{he_0,he_1,he_2\} \in CT_2(TID) $ }
\State $B(1) \Leftarrow he_0 XOR he_1$    \label{alg_line_gpu_h012_check_start}
\State $B(2) \Leftarrow he_0 XOR he_2$
\If{ left $q-\delta$ bits of $B(0)$ not equal to right $q-\delta$ bits of $B(1)$} \label{alg_line_gpu_h012_check_end}
\State Continue
\EndIf                                     
 \State insert $ct$ into $SCTB$
\EndFor
\end{algorithmic}
\end{algorithm}

In algorithm \ref{alg_candidateBuf_H012} it points to $CTB_1$. $CT_2(TID)$ could be acquired from $HSE(0)$, $HSE(1)$, $HSE(2)$ according the index of a GPU thread. When testing candidate tuple in $CTB_2(TID)$, valid candidate tuple which passing checking process from line \ref{alg_line_gpu_h012_check_start} to \ref{alg_line_gpu_h012_check_end} will be stored in $SCTB$ for further checking with hot estimators in other rows.

When all threads finished, $CTB_1$ which has stored all valid candidate tuples extracting from the first three rows will work as reading buffer and the other buffer, $CTB_2$ will be used for storing new candidate tuple as shown in figure \ref{GPU_candidateTupleGrow}.

For $HSE(i)$ where $i \geq 3$, a new candidate tuple for checking is generated from a candidate tuple in reading tuple buffer, candidate tuple buffer which has stored valid candidate tuple, and a hot estimator in it. Then $Q=|RCTB|*|HSE(i)|$ where $|RCTB|$ means the number of candidate tuple storing in reading candidate tuple buffer. When $i$ is an odd number, $RCTB$ points to $CTB_1$, $SCTB$ points to $CTB_2$; when $i$ is an even number, $CBT_1$ and $CBT_2$ exchange roles. A new candidate tuple consists of a hot estimator in $HSE(i)$ and a candidate tuple in $RCTB$. The set of such new candidate tuple to be checked by the $j$th thread, $CT_i(j)$, could be generated from $HSE(i)$ and $RCTB$. Algorithm \ref{alg_updateCandidateTuple} shows how to check new candidate tuples.

\begin{algorithm}                       
\caption{global function on GPU \\ 
         Update candidate tuple}          
\label{alg_updateCandidateTuple}
\begin{algorithmic}                    
\Require {\\  
Row index $i$,\\
Hot estimators set $HSE(i)$,\\
Storing candidate tuple buffer $SCTB$,\\
Reading candidate tuple buffer $RCTB$\\
} 
 
\State $TID \Leftarrow $ thread index 
\State $CT_i(TID) \Leftarrow$ get new candidate tuple from $HSE(i)$ and $RCTB$
 
\For{$ct=\{he_0,he_1,he_2,\cdots,he_{i-1},he_{i}\} \in CT_2(TID) $ }
\State $B(i-1) \Leftarrow he_0 XOR he_1$    
\State $B(i) \Leftarrow he_0 XOR he_2$
\If{ left $q-\delta$ bits of $B(0)$ not equal to right $q-\delta$ bits of $B(1)$} 
\State Continue
\EndIf                                     
 \State insert $ct$ into $SCTB$
\EndFor
\end{algorithmic}
\end{algorithm}

When checking a candidate tuple newly adding a hot estimator in $HSE(i)$, only $B(i-1)$ and $B(i)$ should be tested. After updating candidate tuple with the last row, $SCTB$ contains candidate tuple from which a valid host could be reconstructed. Set $Q=|SCTB|$, $U=Q/V$ and launch $V$ threads. Every thread scans $U$ or $U+1$ reconstructed hosts to estimate their opposite number according their union sliding estimators in the candidate tuple and checks if they are sliding super points. By this method, every thread on GPU has the similar load with the cost of additional buffers for storing middle candidate tuples. Nowadays GPU has plenty global memory and the buffers not occupy many space because the number of sliding super points takes up a small part of hosts. Considering the fast regaining speed, this method is much more suitable for GPU running than the recursive one mentioned before.

\subsection{Distance recorder updation}
After regaining sliding super points at the end of a slot, every distance recorder in sliding estimator should be updated for IP pairs scanning in the next slot. As mentioned before, the updating procedure is very simple, adding by one if the distance recorder does not reach to the maximum. 

There are total $Q=\eta *r*2^q$ distance recorders in $RSEA$. Still use $V$ threads in GPU to update these distance recorders parallel. Let $Q=|SCTB|$, $U=Q/V$ and $W=Q mode V$. Every of the first $W$ threads updates $U+1$ distance recorders and the every of the rest $V-W$ threads updates $U$ distance recorders. 

We denote the GPU version of our sliding super point detection algorithm as GSSD. Experiments on real world core network proves the high accuracy and fast speed of GSSD as shown in next section. 

\section{Experiments}
This paper uses a real world core network traffic downloading from Caida\cite{expdata:Caida}. This traffic contains one-hour IP packets from 13:00 to 14:00 on Febrary 19, 2015. As mentioned before, discrete time window is a special case of sliding window when $k$ is set to 1. Because other algorithms can only work under discrete window, we firstly set $k$ to 1 and $\mu$ to 300 seconds to compare the accuracy and time consumption of our algorithm with that of others. The first discrete window starts from 13:00 and there are total 12 discrete windows in the one-hour IP traffic. Table \ref{tbl-trafficInf} shows the detail information of traffic in these discrete time window.
\begin{table*}
\centering
\caption{Discrete time window traffic information}
\label{tbl-trafficInf}
\begin{tabular}{c}                                                                                                                                                                                                                           
\centering
\includegraphics[width=0.7\textwidth]{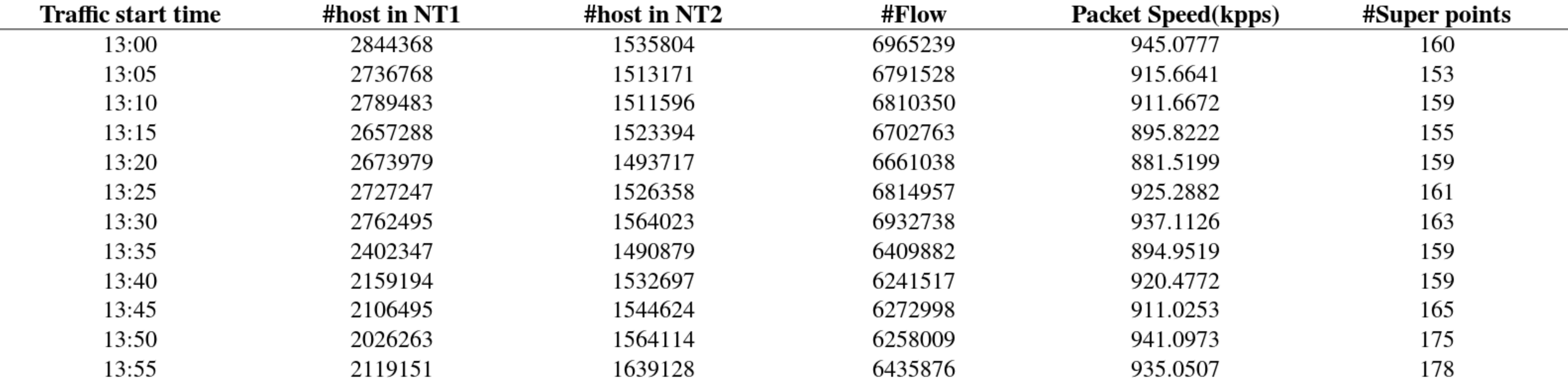}
\end{tabular}
\end{table*}

In our experiments, the threshold $\theta$ of both super point and sliding super point is set to 1024. We implement DCDS \cite{HSD:ADataStreamingMethodMonitorHostConnectionDegreeHighSpeed}, VBFA \cite{HSD:ADataStreamingMethodMonitorHostConnectionDegreeHighSpeed}, GSE \cite{HSD:GPU:2014:AGrandSpreadEstimatorUsingGPU} and our GSSD on a general GPU to evaluate their performance. The GPU card is GTX 950 with 640 CUDA cores and 4 GB of memory and it connects with a PC which has Intel i7 CPU and 8 GB DDR4 memory through PCI-E 3.0 bus. The parameters of GSSD are set as: $\eta = 2^{11}$, $q=14$, $r=5$, $\delta=6$.
IP pairs number $\alpha$ of buffers is set to $2^{15}$. Table \ref{tbl_Avg_dtw_rlt} lists the average detection result of these algorithms on these 12 discrete time windows.
\begin{table*}
\centering
\caption{Super points detection result}
\label{tbl_Avg_dtw_rlt}
\begin{tabular}{c}                                                                                                                                                                                                                           
% \begin{figure}[!ht]
\centering
\includegraphics[width=0.8\textwidth]{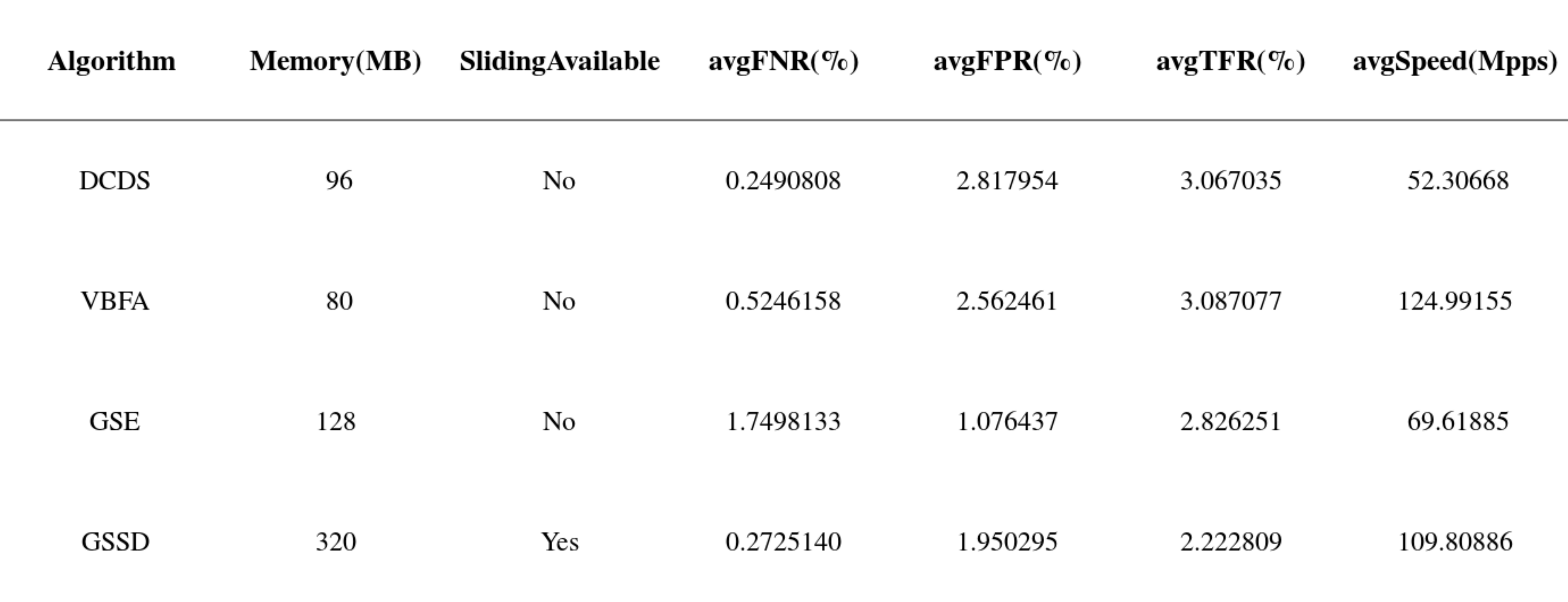}
%\end{figure}
\end{tabular}
\end{table*}

Although GSSD uses the most memory, but only it can detect sliding super point and in the experiment GPU, 4 GB of graphic memory is available for users' program, which is much more than what GSSD requires. 

In our experiments, false positive rate FPR, false negative rate FNR and total false rate TFR are used to evaluate the accuracy of different algorithm. FPR is the ratio of the number of false alarm normal hosts, whose opposite number are smaller than $\theta$ but detected as super points by a algorithm, to the number of super points. FNR is the ratio of the number of super points that fail to be detected to the number of super points. Both FPR and FNR are the smaller the better. But they have the negative correlation: a high FNR will cause a low FPR and vice versa. For an extreme example, FNR could be zero when reports all hosts as super points but FPR will be very big --- the reciprocal of the fraction of super points. So we use TFR, the sum of FPR and FNR, to measure the overall accuracy. 

DCDS has the lowest FNR in these algorithm but its FPR is the highest and it has the lowest speed because it uses CRT to restoring super points. VBFA replaces the CRT in DCDS with bits extraction operation to speed up the packets scanning procedure and it has the fastest speed in all of these algorithm. But bits extraction in VBFA has low randomness which let it has the biggest TFR. Because GSE needs to scan every hosts in a list when regaining super points, so its speed is not very high. But GSE uses a compact structure to estimate hosts opposite number, so its TFR is smaller than that of DCDS and VBFA. GSSD has the highest accuracy in these algorithms because $RSEA$ can estimator host's opposite number more accuracy and $RHFG$ is random enough to make full use of every sliding estimator in $RSEA$. What's more, all procedures, scanning IP pair, regaining super points and updating slots, in GSSD are very simple, so its speed is very high. Supposing the average size of every packet is 800 B, then the throughput of GSSD will reach to 681.25 Gb/s , calculating by $\frac{109*800*8}{1024}$.

\begin{figure*}[!ht]
\centering
\includegraphics[width=0.9\textwidth]{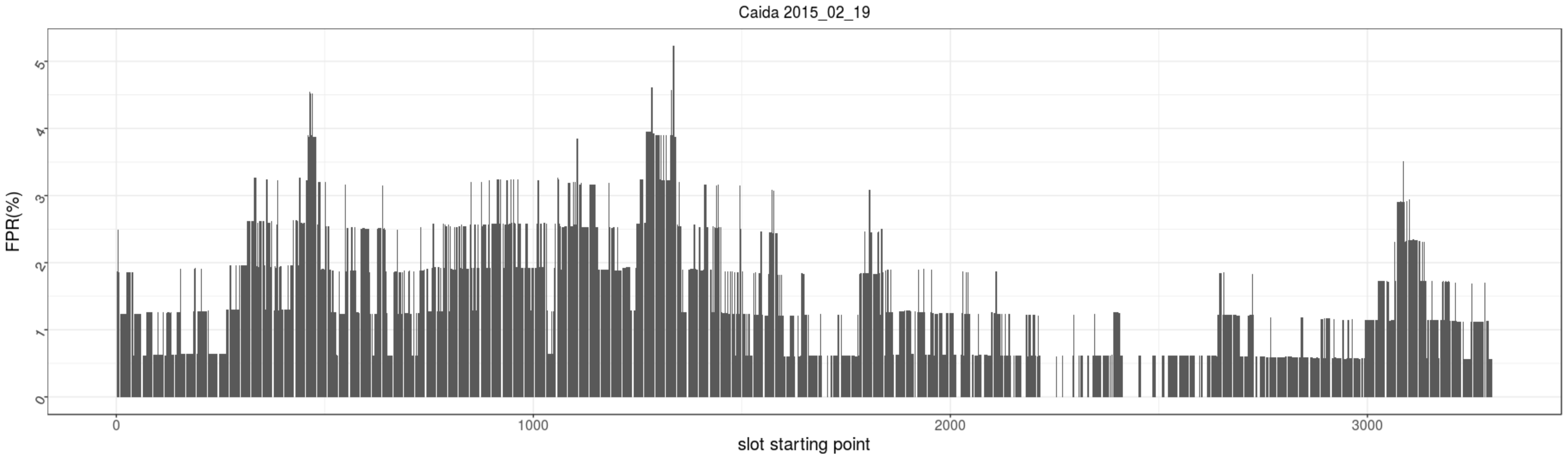}
\caption{FPR under sliding window}
\label{SW_fpr}
\end{figure*}

\begin{figure*}[!ht]
\centering
\includegraphics[width=0.9\textwidth]{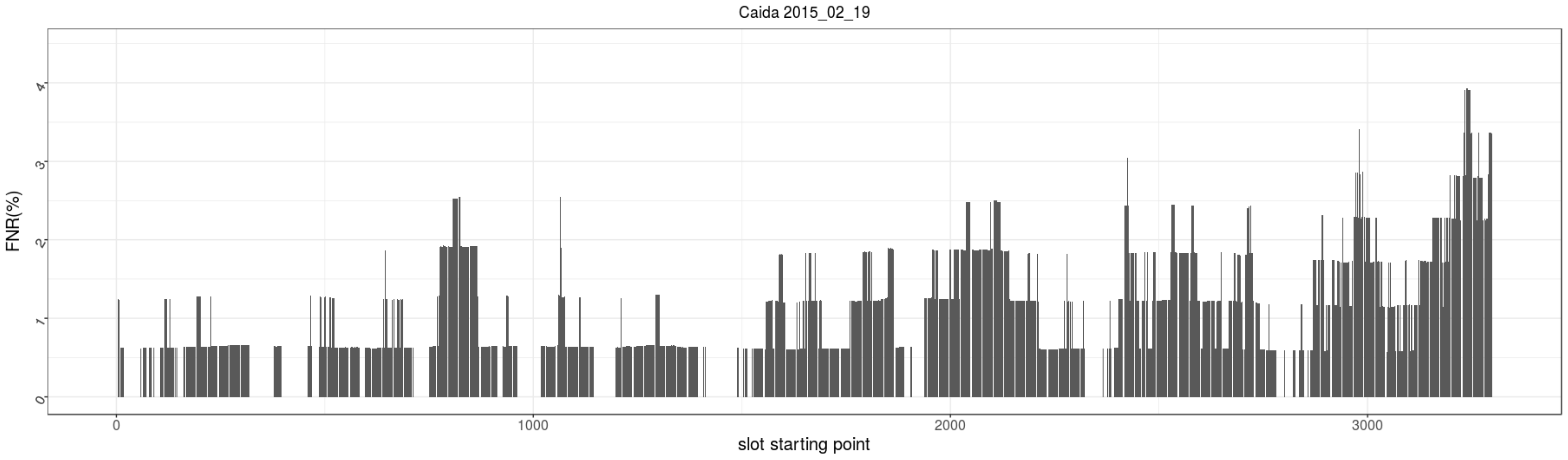}
\caption{FNR under sliding window}
\label{SW_fnr}
\end{figure*}

Not only in discrete window, but also in sliding window GSSD acquires the highest TFR. We set the slot's duration $\mu$ to one seconds and the first slot is the second between 13:00:00 to 13:00:01. There are 3600 slots in the one-hour traffic. In order to have the same time period, k is set to 300. FPR and FNR of GSSD in sliding window, from $SW(0,300)$ to $SW(3299,300)$ are shown in figure \ref{SW_fpr} and \ref{SW_fnr}.

\begin{figure*}[!ht]
\centering
\includegraphics[width=0.9\textwidth]{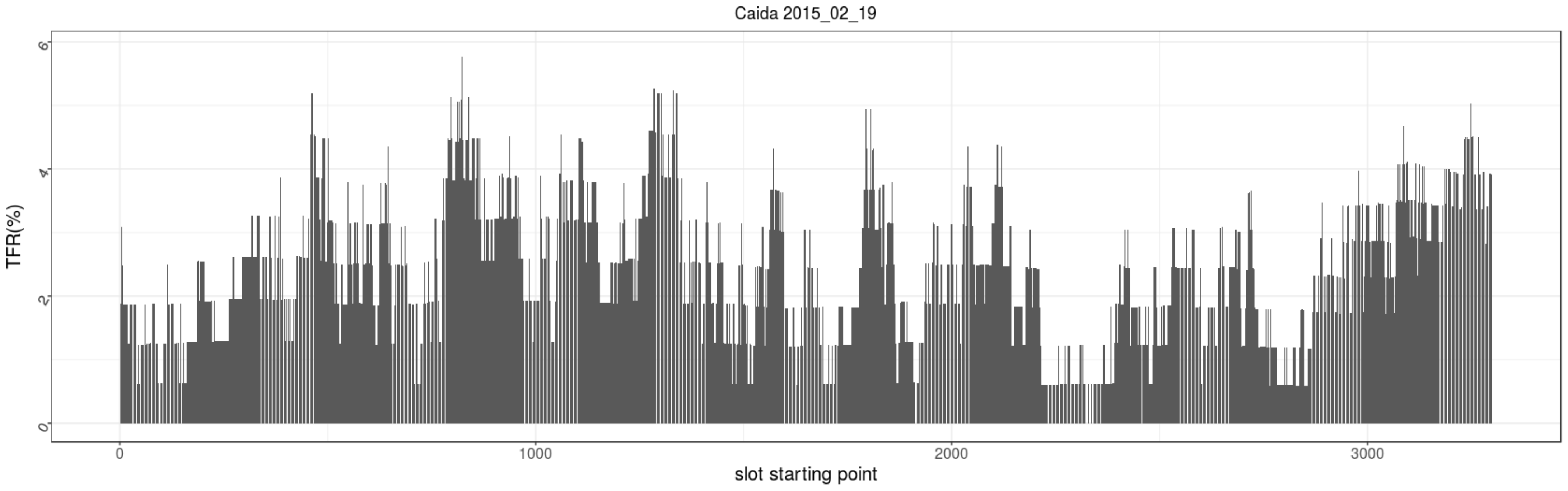}
\caption{TFR under sliding window}
\label{SW_tfr}
\end{figure*}

In most sliding window, GSSD's FPR is smaller than $2\%$ and its average value is $1.43\%$.
FNR of GSSD is much smaller, with average $0.97\%$. Comparing the distribution of FPR and FNR, we can find that when FPR is big, FNR is low and the TFR is tend to $2.4\%$ as shown in figure \ref{SW_tfr}.

This experiments shows that GSSD has low TFR and high speed for sliding super point detection in core network. It can be applied to a bigger network by increasing the sliding estimators in $RSEA$ or using several $RSEA$ to regain different groups of hosts splitting by their IP addresses. 
\section{Conclusion}
Sliding super point measures host's opposite number at a finer level of granularity and won't be affected by the starting of time period. It reports special hosts more timely. Based on the novel proposed sliding estimator array $RSEA$ and hash functions group $RHFG$, this paper firstly devises a sliding super point detection algorithm GSSD for distributed core network. GSSE has simple updating procedure and high detection accuracy. This paper also gives two methods to regain super points from $RSEA$: a recursive version and a parallel version. Recursive method is memory efficient that it only requires a candidate tuple to store intermediate result. But its speed is not very high. Parallel version reconstructs sliding super points incrementally by thousands of threads concomitantly and has the fastest speed at the cost of two additional buffers. GSSD uses parallel hosts regaining method and update $RSEA$ on GPU. When running on a conventional GPU with 640 cores, it can deal with a 680 Gb/s network in real time which is a very big core network. Of course GSSE can deal with a higher speed network, such as Chinese output 7000 Gb/s network, when using one or more advance GPUs on every node. But it still has some detail questions to be solved for such a big network, such as nodes communication, load balance etc. and that's what our future work.

\iftoggle{ACM}{
\bibliographystyle{ACM-Reference-Format}
}

\bibliography{..//..//ref} %replace sigproc with your own bib file

\end{document}